\documentstyle[psfig]{mn}
\def\Mpc{{\rm\thinspace Mpc}}
\def\erg{{\rm\thinspace erg}}
\def\keV{{\rm\thinspace keV}}
\def\km{{\rm\thinspace km}}

\def\s{{\rm\thinspace s}}

\def\keV{{\rm\thinspace keV}}

\def\kmps{\hbox{$\km\s^{-1}\,$}}
\def\kmpspMpc{\hbox{$\kmps\Mpc^{-1}$}}
\def\ergps{\hbox{$\erg\s^{-1}\,$}}
\def\etal{{\it et al.\ }}
\def\eg{{\it e.g.\ }}
\def\apc{\rm atom cm$^{-2}$}
\def\spose#1{\hbox to 0pt{#1\hss}}
\def\approxlt{\mathrel{\spose{\lower 3pt\hbox{$\sim$}}        \raise 2.0pt\hbox{$<$}}}
\def\approxgt{\mathrel{\spose{\lower 3pt\hbox{$\sim$}}        \raise 2.0pt\hbox{$>$}}}

\begin{document}

\title{The ASCA X-ray spectrum of the powerful radio galaxy 3C109}

\author[]
{\parbox[]{6.in} {S.W. Allen$^1$, A.C. Fabian$^1$, 
E. Idesawa$^2$, H. Inoue$^3$, T. Kii$^3$, C. Otani$^4$ \\
\footnotesize
1. Institute of Astronomy, Madingley Road, Cambridge CB3 0HA,  \\
2. Department of Physics, University of Tokyo, Hongo, Bunkyo-ku, Tokyo,
Japan \\
3. Institute of Space and Astronautical Science, Yoshinodai, Sagamihara,
Kanagawa 229, Japan \\
4. RIKEN, Institute of Physical and Chemical Research, Hirosawa, Wako,
Saitama 351-01, Japan \\
}}
\maketitle

\begin{abstract} We report the results from  an ASCA X-ray observation of the
powerful Broad Line Radio Galaxy, 3C109. The ASCA 
spectra confirm our earlier ROSAT
detection of intrinsic X-ray absorption 
associated with the source. The absorbing
material obscures a 
central engine of quasar-like luminosity. The luminosity is 
variable, having dropped 
by a factor of two since the ROSAT observations 4
years before. 
The ASCA data also provide evidence for a broad iron
emission line from the source, with an intrinsic FWHM of 
$\sim 120,000\kmps$. Interpreting the line as
fluorescent emission from the inner parts of an accretion disk, we can
constrain the inclination of the disk to be $> 35$ degree, and the inner
radius of the disk to be $< 70$ Schwarzschild radii. Our results
support unified schemes for active galaxies, and demonstrate a remarkable 
similarity between the X-ray properties of this powerful radio source, and 
those of lower luminosity,  Seyfert 1 galaxies.
\end{abstract}


\section{Introduction} 

Unified models of radio sources propose that radio galaxies and radio-loud
quasars are basically the same population of objects, viewed at different
orientations (Orr \& Browne 1982; Scheuer 1987; Barthel 1989). The nucleus is
only directly visible in quasars, the radio axis of which points within
$\sim45$ degree of the line of sight. In the case of radio galaxies the axis is
closer to the plane of the Sky and the nucleus is obscured from view by
material in the host galaxy, possibly in a toroidal distribution.

The powerful Broad Line Radio Galaxy (BLRG) 3C109 appears to be 
oriented at an intermediate
angle. The nucleus is reddened, $E(B-V) \sim 0.9$, 
and polarized in the optical waveband
(Rudy \etal 1984; Goodrich
\& Cohen 1992), suggesting that our line of sight passes through the edge 
of the obscuring material.
The dereddened luminosity of the nucleus,  $V=-26.2$ (Goodrich \& Cohen 1992)
identifies the source as an intrinsically luminous
quasar. Obscuration is also seen at X-ray wavelengths (Allen \& Fabian 1992).
3C109 was serendipitously observed 
with the Position Sensitive Proportional Counter (PSPC) on ROSAT 
in 1991 August. The PSPC  
spectrum exhibits soft X-ray absorption in excess 
of that expected from material within our own Galaxy, implying  an intrinsic
equivalent hydrogen column
density at the redshift of the source ($z = 0.3056$; Spinrad \etal 1985) 
of $\sim 5 \times 10^{21}$\apc. The intrinsic (unabsorbed)
X-ray luminosity of the source ($0.1-2.4$ keV) 
determined from the PSPC data is $\sim 5\times 10^{45}$\ergps,  
 making it one of the most X-ray
luminous objects within $z\sim 0.5$; only the QSOs 3C273 and E1821+643
have higher X-ray luminosities (and 3C273 may have a
significant beamed component to its X-ray emission). 

We present here the results of an ASCA X-ray observation of 3C109. The
ASCA data confirm the results of Allen \& Fabian (1992) on excess 
absorption, and allow us to
explore further the X-ray properties of this remarkable source. 
We show that 3C109 has decreased in brightness by
about a factor of two since the ROSAT observations,  to a flux 
level comparable with that observed with the Imaging Proportional Counter
(IPC) on the {\it Einstein Observatory} in 1979
(Fabbiano et al 1984). Also, of particular interest 
is the detection of a strong, broad iron
line in the ASCA spectra. This result implies that most of the X-ray emission 
from 3C109  is unbeamed. Modelling the line as fluorescent Fe K 
emission from an accretion disk, we are able to constrain both the 
inclination and inner radius of the disk. The X-ray properties of 3C109
are shown to be remarkably similar to those of many lower-power, 
Seyfert 1 galaxies. 
Throughout this paper we assume a value for the Hubble constant of
$H_0$=50 \kmpspMpc and a cosmological deceleration parameter $q_0$=0.5.

\section{The ASCA observations}

The ASCA X-ray astronomy satellite (Tanaka, Inoue \& Holt 1994) 
consists of four separate nested-foil telescopes, each with 
a dedicated X-ray detector. 
The detectors include two  Solid-state Imaging
Spectrometers or SISs (Burke \etal 1991, Gendreau 1995) 
and two Gas scintillation Imaging Spectrometers or GISs (Kohmura \etal 1993).
The SIS instruments provide 
high quantum efficiency and good spectral resolution, $\Delta E/E =
0.02(E/5.9 {\rm keV})^{-0.5}$. The GIS detectors provide a lower resolution, 
$\Delta E/E =  0.08(E/5.9
{\rm keV})^{-0.5}$, but cover a larger 
($\sim 50$ arcmin
diameter) circular field of view.

3C109 was observed with ASCA on 1995 Aug 28-29.
The SIS observations were made in the standard
1-CCD mode (Day \etal 1995)  with the source positioned  at 
the nominal pointing position for 
 this mode. X-ray event lists were constructed using 
the standard screening criteria and data reduction techniques discussed 
by Day \etal (1995).
The observations are summarized in Table 1. 

Source spectra were extracted from circular regions of  
radius 4 arcmin (SIS0), 3.5 arcmin (SIS1) and 6 arcmin (GIS2, GIS3), 
respectively. For the SIS data, background spectra were 
extracted from regions of
the chip relatively free of source counts. 
For the GIS data, background
spectra were extracted from circular regions, the same size as the
source regions, and at similar 
distance from the optical axes of the telescopes. 

Spectral analysis was carried out using the 
XSPEC spectral
fitting package (Shafer \etal 1991). For the SIS data, the 1994 Nov 9
version of the SIS response matrices were used. 
For the GIS data the 1995 Mar 6 response matrices were used. 
The spectra were binned to have a minimum of 20 counts per Pulse Height
Analysis (PHA) channel, thereby allowing $\chi^2$ statistics to be used. 
In general, best-fit parameter values and confidence limits quoted in the 
text are the results from simultaneous fits 
to all 4 ASCA data sets, with the normalization of the 
power-law continuum allowed to vary independently for each data set. 

\begin{table*}
\vskip 0.2truein
\begin{center}
\caption{Observation summary}
\vskip 0.2truein
\begin{tabular}{ c c c c c }

\hline
Instrument    & ~ & Observation Date   & ~ &     Exposure (ks)   \\

&&&& \\
ASCA SIS0      & ~ & 1995 Aug 28/29     & ~ &          36.0              \\
ASCA SIS1      & ~ &    ""              & ~ &          35.0              \\
ASCA GIS2      & ~ &    ""              & ~ &          35.0              \\
ASCA GIS3      & ~ &    ""              & ~ &          35.0              \\
&&&& \\
ROSAT PSPC     & ~ & 1991 Aug 30        & ~ &          22.1              \\
Einstein IPC   & ~ & 1979 Mar 7         & ~ &          1.86              \\
\hline
&&&& \\
\end{tabular}
\parbox {4.5in}
{ Notes: X-ray observations of 3C109. Exposure
times are for the final X-ray event lists after standard screening
criteria and corrections have been applied. 
}
\end{center}

\end{table*}

\section{Results}

\subsection{Confirmation of excess X-ray absorption in 3C109}

\begin{figure}
\centerline{\hspace{3cm}\psfig{figure=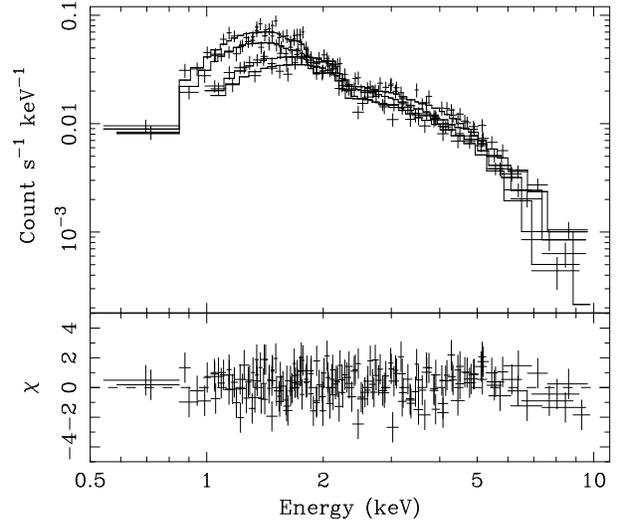,width=0.65\textwidth,angle=270}}
\caption{(Upper panel) The SIS and GIS spectra of 3C109 with the best
fitting absorbed power-law model (Model A) overlaid. (Lower panel) 
The residuals to the fit in units of $\chi$. (For plotting purposes the
data have been rebinned along the energy axis by a factor 7.)}
\end{figure}

\begin{figure}
\centerline{\hspace{3cm}\psfig{figure=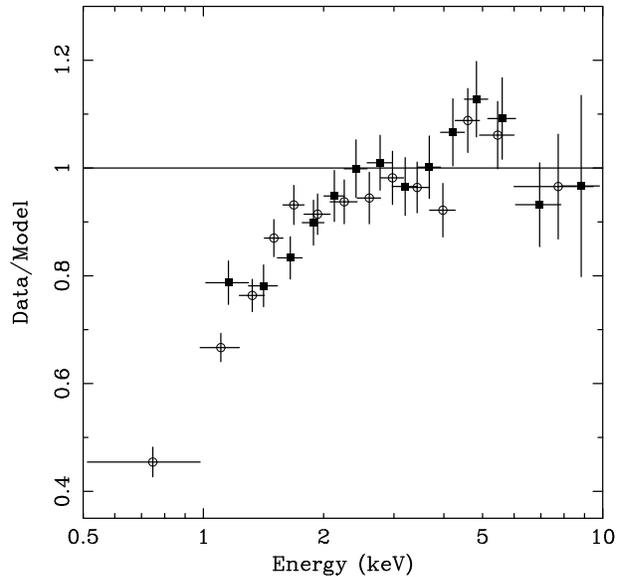,width=0.65\textwidth,angle=270}}
\caption{(Upper panel) The ratio of data to model, where the
model is the best-fit Model A, but with the absorption reset 
to the Galactic value
(assumed to be $3 \times 10^{21}$ \apc). Note 
the large negative residuals at energies below 2 keV which are due to the
excess absorption, and the evidence for a broad, redshifted emission line feature at 
$\sim 5\keV$. For plotting purposes, the SIS 
(open circles) and GIS (filled squares) data sets have been
averaged together and binned by a factor of 20 along the energy axis.}
\end{figure}

\begin{figure}
\centerline{\hspace{3cm}\psfig{figure=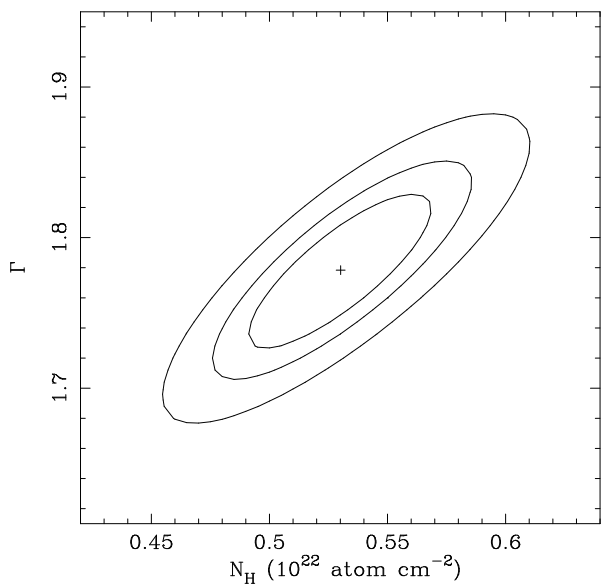,width=0.65\textwidth
,angle=270}}
\caption{Joint confidence contours on the photon index and total column
density, determined with spectral Model A (Table 3). 
Contours mark the regions of 68, 90 and 99
per
cent confidence ($\Delta \chi^2 = 2.30, 4.61$ and 9.21 respectively).
}
\end{figure}

\begin{figure}
\centerline{\hspace{3cm}\psfig{figure=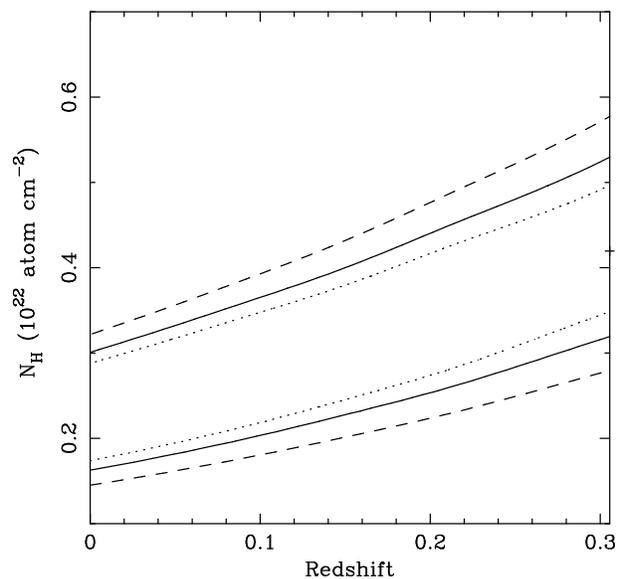,width=0.65\textwidth,angle=270}}
\caption{Joint confidence contours (68, 90 and 99
per cent confidence) on the column density and redshift of the 
excess absorber in 3C109 (using spectral Model B).}
\end{figure}

The principal result of the ROSAT PSPC observation of 3C109 (Allen \&
Fabian 1992) was the
detection of X-ray absorption in excess of the Galactic value determined
from 21 cm HI observations. 
The ASCA data allow us to verify and expand upon this result. 

The ASCA data were first examined using a simple absorbed power law model.  
This
allows direct comparison with the results of Allen \& Fabian (1992). The
free parameters in the fits were the column density of the absorbing
material, 
$N_{\rm H}$,  the photon index 
of the power law emission, $\Gamma$,  (both parameters were forced to take 
the same value in
all 4 ASCA data sets) and the normalizations, $A_1$, of the power-law
emission. 
(Due to the range of source extraction regions used, 
and known systematic differences in the flux calibration of the 
different ASCA detectors, the value of $A_1$ 
was allowed to vary independently for each data set). 
The best
fit parameter values and 90 per cent ($\Delta \chi^2 = 2.71$) 
confidence limits obtained with this simple model
are presented in Table 3 (Model A). 

The SIS and GIS spectra
with their best-fitting models (Model A) overlaid are plotted in Fig. 1.
For illustrative purposes, in Fig. 2 we show the best fit model
with the column density reset to the Galactic value (assumed to be 
$3.0 \times 10^{21}$ \apc). 
Note the large negative residuals at energies, $E < 2$ keV, which 
demonstrate the effects of the excess absorption, and the broad 
positive residual at $E \sim 5$ 
keV, which will be discussed in more detail in Section 3.3. 
  
The ASCA results clearly confirm the PSPC result on excess 
absorption in the X-ray spectrum of 3C109. Assuming that the absorber lies
at zero redshift we determine a total column density along the line of
sight of $5.30\pm0.42
\times 10^{21}$ \apc (90 per cent confidence limits). 
This is in good agreement with 
the PSPC result of $4.2^{+1.9}_{-1.6} \times 
10^{21}$ \apc. The ASCA result on the 
photon index, $\Gamma = 1.78^{+0.05}_{-0.06}$, is also in excellent agreement 
with the PSPC result of $1.78^{+0.85}_{-0.76}$, although is more
firmly constrained.
The joint confidence contours on $\Gamma$ and 
$N_{\rm H}$ are plotted in Fig. 3.

We have examined the constraints the 
ASCA spectra can place on the redshift of the excess 
absorbing material. 
The Galactic column density along the line of sight to 
3C109, determined from  21cm observations,  is 
$1.46 \times 10^{21}$ \apc (Jahoda \etal
1985; Stark \etal 1992),  although Johnstone \etal (1992) suggest a 
slightly higher value of $\sim 2.0 \times 10^{21}$ \apc, and Allen \&
Fabian (1996) infer a value of $\sim 3.0 \times 10^{21}$ \apc~from X-ray
studies of the nearby cluster of galaxies Abell 478.
Modelling the ASCA spectra with a two-component absorber, with 
a Galactic (zero-redshift) column density of $3.0 \times 10^{21}$ \apc, 
and a component with variable column density and redshift, we obtain the 
joint confidence contours on the redshift and column density of the excess 
absorption plotted in Fig. 4. The best-fit parameter
values and 90 per cent confidence limits for the two-component absorption
model (Model B) are also summarized in Table 3. 

\subsection{Variation of the X-ray luminosity}

\begin{table*}
\vskip 0.2truein
\begin{center}
\caption{X-ray flux of 3C109}
\vskip 0.2truein
\begin{tabular}{ c c c c c c c c c }
\hline
\multicolumn{1}{c}{Instrument} &
\multicolumn{1}{c}{} &
\multicolumn{1}{c}{Date} &
\multicolumn{1}{c}{} &
\multicolumn{2}{c}{$F_X$} &
\multicolumn{1}{c}{} &
\multicolumn{2}{c}{$L_X$}  \\
&&&& \\
             & ~ &                & ~ &  2-10 keV     & 1-2 keV       & ~ & 2-10 keV     & 1-2 keV       \\
ASCA SIS0    & ~ & 1995 Aug 28/29 & ~ & $48.5\pm1.6$  & $9.50\pm0.23$ & ~ & $21.4\pm0.3$ & $6.69^{+0.76}_{-0.71}$ \\
ASCA SIS1    & ~ &   ""           & ~ & $46.3\pm2.1$  & $9.76\pm0.32$ & ~ & $21.3\pm0.3$ & $7.68^{+1.10}_{-0.89}$ \\
&&&& \\
             & ~ &                & ~ & 0.1-2.4 keV   & 1-2 keV       & ~ & 0.1-2.4 keV  & 1-2 keV       \\
ROSAT PSPC   & ~ & 1991 Aug 30    & ~ & $28.9\pm2.6$  & $18.2\pm0.7$  & ~ & $45^{+147}_{-25}$
& $12.1^{+6.3}_{-3.5}$  \\
&&&& \\
             & ~ &                & ~ &  0.5-3.0 keV  & 1-2 keV       & ~ & 0.5-3.0 keV  & 1-2 keV       \\
Einstein IPC & ~ & 1979 Mar 7     & ~ & $20\pm6$   & $8.3\pm2.5$   & ~ & $16.5\pm5.0$ & $6.2\pm1.9$  \\
\hline
&&&& \\
\end{tabular}
\parbox {7in}
{ Notes: The X-ray flux and luminosity of 3C109 measured
with ASCA, ROSAT and the Einstein Observatory. Fluxes 
are in units of $10^{-13}$ erg cm$^{-2}$s$^{-1}$ 
and are defined in the rest frame of the
observer.  
Luminosities are in $10^{44}$ erg s$^{-1}$, are absorption corrected, 
 and are quoted in the 
rest frame of the source. 
Errors are 90 per cent 
($\Delta \chi^2 = 2.71$) confidence limits.}
\end{center}
\end{table*}

The flux measurements for 3C109 are summarized in Table 2. Results
are presented for both SIS instruments in the $1.0-2.0$ and
$2.0 - 10.0$ keV (observer frame) energy bands. (The GIS detectors provide
less accurate flux estimates). Also listed in Table 2 are the fluxes
observed with the ROSAT PSPC in August 1991 and the IPC on 
Einstein Observatory in March 1979. We see
that in the overlapping $1.0-2.0$ keV energy band, the brightness of 3C109 
has decreased by a factor $\sim 2$ since 1991. The flux determination 
from 
the ASCA data is now consistent with that inferred from the IPC observation
in 1979. 

Also listed in Table 2 are the intrinsic (absorption-corrected)
X-ray luminosities of the source inferred from the observations. 
(Here the energy bands correspond to the rest-frame of the source). 
The absorption-corrected $2-10$ keV luminosity inferred from the ASCA spectra
is $2.1 \times 10^{45}$ \ergps. (We assume that during the Einstein IPC
observations the source had the same 
spectral shape as determined from the
ASCA observations.)
 
3C109 has also been observed to
vary at near-infrared wavelengths. Rudy \etal (1984) found variations
of a factor $\sim 2$ in the $J$ band over a five year span from 1978 to
1983. Elvis \etal (1984) similarly reported variations in the  $J,H$ and $K$
bands of $\sim 50$ per cent (in the same sense) on a timescale of 2--3 years
between 1980 and 1983.

\subsection{Discovery of a broad iron line}

The residuals to the fits with the simple power-law models, presented 
in Figs. 1 and 2, exhibit an excess of counts in a line-like feature 
at $E \sim 5.0$ keV. X-ray observations of Seyfert
galaxies (Nandra \& Pounds 1994 and references therein) 
show that many such sources exhibit a strong emission
line at $E \sim 6.40$ keV (in the rest frame of the object). This is normally 
attributed to fluorescent Fe K emission from cold material irradiated by
the nucleus. 

We find that the fit to the ASCA data for 3C109 is significantly improved by 
the introduction of a Gaussian line at $E \sim 5$ keV 
($\Delta \chi^2 = 9.2$ for 3 extra
fit parameters; an F-test indicates this to be significant at
the 97 per cent level.) The best-fit line energy is 
$5.09^{+0.44}_{-0.38}$ keV (corresponding to $6.61^{+0.57}_{-0.50}$ keV
in the rest frame of the source.  Note that if a fixed rest-energy of
6.4 keV is assumed, the introduction of the Gaussian component 
becomes significant at the $\sim 99$ per cent confidence level). 
The data also indicate that 
the line is broad, with a 1 sigma width of $0.65^{+0.81}_{-0.36}$ keV. The
equivalent width of the line is $300^{+600}_{-200}$ eV.  
The width and energy of the line suggest that it is due to fluorescence
from a rapidly rotating accretion disk -- as is thought to be the case 
in lower luminosity
Seyfert galaxies (Tanaka \etal 1995; Fabian \etal 1995).    
The best fitting parameters and confidence limits for the power-law plus 
Gaussian model (Model C) 
are summarized in Table 3. 
Note that the emission feature is not well-modelled by the introduction
of an absorption edge at higher energies [the introduction of an edge into the simple
absorbed power-law model (Model A) does not significantly improve the fit]. 
Note also that the measured line flux is not 
significantly affected by the small systematic bump in the XRT response at 
$E \sim 5.5$ keV (which produces a narrow, positive residual with a flux
of a few per cent of the continuum flux at that energy).

\subsection{Modelling the line as a diskline}

\begin{figure}
\centerline{\hspace{3.0cm}\psfig{figure=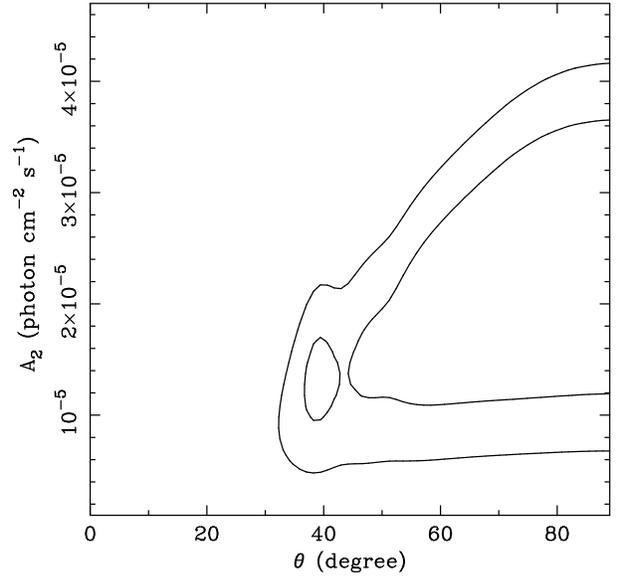,width=0.65\textwidth
,angle=270}}
\caption{Joint confidence contours (68, and 90
per cent confidence) on the normalization, $A_2$, and inclination,
$\theta$, of the
disk line using spectral Model D (following Fabian \etal 1989).}
\end{figure}

Although the simple Gaussian model provides a reasonable description of the
5.0 keV emission feature, the ASCA data suggest that the line profile 
is probably more complex. Using two Gaussian components to model the line
profile, we obtain the best fit for a broad component with
a rest energy consistent with 6.4 keV, and a 
narrow component with an energy $6.8 \pm 0.1$ keV (in the rest-frame of the 
source). These results are similar to those obtained
for nearby, lower-luminosity Seyfert galaxies (\eg Mushotzky et al 1995; 
Tanaka et al 1995; Iwasawa \etal 1996) where the line emission is 
thought to originate from the inner regions of an accretion disk surrounding
a central, massive black hole (Fabian \etal 1989). 

We have therefore modelled the broad line in 3C109 using the Fabian \etal 
(1989) model for line emission from a relativistic accretion disk.
The rest-energy of the line (in the emitted frame)  was fixed at 6.40 
keV, the energy appropriate for Fe K fluorescence from cold material.
(The effects of cosmological redshift were incorporated into the model.)
 The accretion 
disk was assumed to extend over radii from 3 to 500 Schwarzschild radii
(hereafter $R_{\rm s}$)
and cover a solid angle of $2\pi$ steradians. The emissivity was assumed
to follow a standard disk radiation law.
Only the disk inclination and line strength were free parameters in the fit.
The best fit parameters and 90 per cent confidence limits obtained with
the diskline model are listed in
Table 3 (Model D). 
The introduction of the diskline component significantly
improves the fit to the ASCA data with respect to 
the power-law model ($\Delta \chi^2 = 8.9$ for 2 extra fit
parameters, which  an F-test indicates to be significant at the 
99 per cent confidence level).

In Fig. 5 we show the joint confidence contours 
on the inclination of the disk, $\theta$, versus the line strength, $A_2$.
The 90 per cent ($\Delta \chi^2 =
2.71$) constraint on the inclination is $\theta > 35$ degree.  
We have also examined the constraints that may be placed on the 
inner radius, $r_{in}$, of the accretion
disk with the diskline model. The data were re-fitted with $r_{in}$
included as a free parameter. The preferred value for $r_{in}$  is
3 $R_{\rm s}$, with a 90 per cent confidence upper limit of
$70 R_{\rm s}$.  (Note that for an ionized disk, with a rest-energy for
the line of 6.7 keV, the inclination is constrained to 
$\theta > 18$ degree.)

The effects of introducing a further, flatter
power-law component into the fits, such as may be required to
account for reflected emission from the illuminated face of an 
accretion disk, or 
synchrotron self-Compton emission from within a jet, were also examined.
The introduction of a flatter power-law component does not significantly 
improve the fits. However, the ASCA spectra permit (with no significant 
change in $\chi^2$) the inclusion of a continuum
spectrum appropriate for reflection from a cold disk, subtending a solid angle
of $2\pi$ steradians to the primary X-ray source, oriented at any
inclination consistent with the results from the diskline fits.

\begin{table*}
\vskip 0.2truein
\begin{center}
\caption{Results of the spectral analysis}
\vskip 0.2truein
\begin{tabular}{ c c c c c c c c c }
\hline
\multicolumn{1}{c}{MODEL} &
\multicolumn{1}{c}{} &
\multicolumn{6}{c}{PARAMETERS} &
\multicolumn{1}{c}{} \\
&&&&&&&& \\                                                                                                

A                & ~ &      $\Gamma$              &           $A_1$         & $N_{\rm H}$               & --- & --- & --- &   $\chi^2$/DOF     \\ 
wabs (pow)       & ~ & $1.78^{+0.05}_{-0.06}$     &  $1.38^{+0.10}_{-0.10}$ & $0.530^{+0.042}_{-0.042}$ & --- & --- & --- & 638.9/633  \\
&&&&&&&& \\                                                                                                
B                & ~ &      $\Gamma$              &           $A_1$         & $N_{\rm H}$               & $N_{\rm H}(z)$            & --- & --- & $\chi^2$/DOF      \\
wabs zwabs (pow) & ~ & $1.77^{+0.05}_{-0.06}$     &  $1.35^{+0.10}_{-0.09}$ & $0.300$                   & $0.420^{+0.083}_{-0.078}$ & --- & --- & 641.6/633   \\                
&&&&&&&& \\                                                                                                
C                & ~ &      $\Gamma$              &          $A_1$          & $N_{\rm H}$               &  $E$                   & $\sigma$                                   &  $A_2$                  &  $\chi^2$/DOF     \\              
wabs (pow+gau)   & ~ & $1.86^{+0.12}_{-0.08}$    &
$1.47^{+0.16}_{-0.12}$ & $0.558^{+0.056}_{-0.046}$ & $5.09^{+0.44}_{-0.38}$ & $0.65^{+0.81}_{-0.36}$                  &  $2.1^{+4.3}_{-1.3}$ & 629.7/630  \\
&&&&&&&& \\                                                                                                
%D                & ~ &   $\Gamma$              &          $A_1$           & $N_{\rm H}$               &  $E_1$                   & 
%$E_2$                                          &  $\sigma_2$              &   $\chi^2$/DOF     \\
%\hline
%wabs (pow+2gau)  & ~ & $1.85^{+0.018}_{-0.08}$ & $1.43^{+0.27}_{-0.08}$   & $0.554^{+0.075}_{-0.043}$ & $6.79^{+0.11}_{-0.10}$   & 
%$6.40$                                         &  $0.97^{+2.61}_{-0.80}$  & 627.3/629  \\
%&&&&&&&& \\                                                                                                
D                & ~ &   $\Gamma$              &          $A_1$         & $N_{\rm H}$               &  $E$                   & $\theta$                  
                 &  $A_2$                  &   $\chi^2$/DOF     \\
wabs (pow+diskline) & ~ & $1.87^{+0.08}_{-0.08}$     &
$1.48^{+0.14}_{-0.13}$ & $0.561^{+0.052}_{-0.046}$ & $6.40$ & $90^{+0.0}_{-55}$                  &  $2.4^{+1.4}_{-1.4}$ & 630.0/631  \\
&&&&&&&& \\                                                                                                
\hline
&&&&&&&& \\                                                                                                

\end{tabular}
\parbox {7in}
{ Notes:  A summary of best-fit parameters and 90 per cent
($\Delta \chi^2 = 2.71$) 
confidence limits from the spectral analysis of the ASCA data. Results are shown for four different models 
fitted simultaneously  to the data for all four ASCA detectors. 
$\Gamma$ is the photon index of the underlying power-law continuum from the source. $A_1$ is the 
normalization of the power law component in the S0 detector in $10^{-3}$ photon
 keV$^{-1}$cm$^{-2}$s$^{-1}$ at 1 keV. $N_{\rm H}$ is the equivalent 
hydrogen column density in $10^{22}$ \apc at zero redshift. In Model B, $N_{\rm H}$(z) is the best-fit intrinsic column density  
at the source for an assumed Galactic column density of $0.3 \times 10^{22}$ \apc.  
In Model C, $E$ is the energy of the Gaussian emission line in
the frame of the observer, $\sigma$ is the one-sigma line width in keV,  and $A_2$ is the line strength in 
$10^{-5}$photon cm$^{-2}$s$^{-1}$. In Model 
D, $E$ is the rest-energy of the line in the emitted frame, $\theta$ is
the inclination of the disk in degree, and $A_2$ is again 
the line strength in $10^{-5}$photon cm$^{-2}$s$^{-1}$. 
}
\end{center}
\end{table*}

\section{Discussion}

The ASCA results on excess X-ray absorption in 3C109 confirm and refine the
earlier ROSAT results (Allen \& Fabian 1992). The ASCA data show (under
the assumption that all of the absorbing material lies at zero redshift) 
that the X-ray spectrum of the source is absorbed by a total column density 
of $5.30^{+0.42}_{-0.42} \times 10^{21}$ \apc~(Model A). 
This compares to a Galactic
column density of $\sim 3.0 \times 10^{21}$ \apc (Allen \& Fabian 1996). 
If we instead assume that the excess absorption, over and above the Galactic
value, is due to material at the
redshift of 3C109, we determine an intrinsic column density of 
$4.20^{+0.83}_{-0.78} \times 10^{21}$ \apc. Note that these 
results assume solar abundances in the absorbing material (Morrison 
\& McCammon 1983).

The X-ray absorption measurements are in good agreement with 
optical results on the polarization and intrinsic reddening of the 
source. Goodrich \& Cohen (1992) determine an intrinsic continuum 
reddening of $E(B-V) \sim 0.9$, 
in addition to an assumed Galactic reddening of $E(B-V) = 0.27$. 
Using the standard (Galactic) relationship between $E(B-V)$ and X-ray
column density,  $N_{\rm H}$/$E(B-V) = 5.8
\times 10^{21}$ atom cm$^{-2}$ mag$^{-1}$ (Bohlin, Savage \& Drake 1978),
the total reddening observed, $E(B-V) \sim 1.2$, 
implies a total X-ray column density (Galactic plus  intrinsic) 
of $\sim 7.0 \times 10^{21}$ \apc.
This result is similar to the X-ray column density inferred from the ASCA
spectra using model B and confirms the presence of significant 
intrinsic absorption at the source.  Note that this result also suggests
that the dust-to-gas ratio in 3C109 is
similar to that in our own Galaxy.

Further constraints on the distribution of the absorbing gas are
obtained from the optical emission-line data presented by Goodrich \&
Cohen (1992). In the narrow line region (NLR), the observed Blamer decrement 
of H$\alpha$/H$\beta = 5.8$ implies 
(for an assumed recombination ratio of 3.2) an $E(B-V)$ value $\sim 0.48$.
Using the relationship of Bohlin, Savage \& Drake
(1978) this implies an X-ray column density to the NLR of $\sim 2.8 \times
10^{21}$ \apc, in good agreement with the Galactic
column density of $\sim 3.0 \times 10^{21}$ \apc~determined  by Allen \&
Fabian (1996) and adopted in the X-ray analysis presented here.  
The Balmer decrement in the broad line
region (BLR) is very steep (H$\alpha$/H$\beta = 13.2$). Although this
value cannot be reliably used to infer the  extinction to the
BLR, the intrinsic line ratio is unlikely to be above 5, suggesting
a total line-of sight reddening to the BLR of $E(B-V) \approxgt
0.8$. Thus, the BLR is likely to be intrinsically reddened by
$E(B-V) \approxgt 0.3$. The optical emission line results are
therefore consistent with the two-component absorber
model (B), with the column density of the intrinsic absorber being 
comparable with the Galactic component.

3C109 is the most powerful object in which 
a strong broad iron line has been resolved to date.
Several more luminous quasars observed with ASCA 
do not show any iron emission or reflection
features (Nandra et al 1995). The next most luminous
object with a confirmed broad line is 3C390.3 (Eracleous, Halpern \&
Livio 1996) which is about 10 times less luminous in both the 
X-ray and radio bands than 3C109. The equivalent widths of the lines in 
both objects are $\sim 300$ eV and therefore similar to those observed in 
lower-luminosity Seyferts. This argues against any X-ray
`Baldwin effect' (as proposed by Iwasawa \& Taniguchi 1993).

The line
emission from 3C109 is most
plausibly due to fluorescence from the innermost regions of an accretion disc
around a central black hole (Fabian et al 1995). Our results constrain the 
inner radius of the accretion disk to be $< 70 R_{\rm s}$ and the inclination
of the disk to be $ > 35$ degree.
The strong iron line observed in 3C109,  and the 
lack of evidence for a synchrotron self-Compton continuum in the X-ray
spectrum, both suggest that little radiation from the jet is beamed
into our line of sight.

The inclination determined from the ASCA data is larger than the 
angle proposed by 
Giovannini \etal (1994) based on the jet/core flux ratio of the source
($\theta < 34$ degree). 
However, the jet/core flux arguments are 
based on simple assumptions about the average orientation
angles for radio galaxies and neglect environmental effects. 
The conflict with the X-ray results may indicate that the 
situation is more complicated. Giovannini \etal (1994) also present
constraints on the inclination from VLBI observations of the 
jet/counterjet ratio, which require $\theta < 56$ degree.
The VLBI constraint, together with the ASCA X-ray constraint, then 
suggests $35 < \theta < 56$ degree.

Our results on 3C109 are in good agreement with the unification schemes 
for radio sources and illustrate the power of X-ray observations for
examining such models. The preferred, intermediate inclination angle for 
the disk in 3C109 is in good agreement with the results on X-ray 
absorption, polarization and optical reddening of the source, all of which 
suggest that 
our line of sight to the nucleus passes closes to the edge of the 
surrounding molecular torus.
The results on the broad iron line reveal a striking similarity
between the X-ray
properties of 3C109 and those of lower 
power, Seyfert 1 galaxies (Mushotzky et al 1995; Tanaka et al 1995; 
Iwasawa et al 1996). This is despite the fact that the X-ray power of 3C109 
exceeds that of a typical Seyfert galaxy by $\sim 2$ orders of magnitude.

\section{ACKNOWLEDGEMENTS} We thank K. Iwasawa, C. Reynolds and R. 
Johnstone for discussions and the annonymous referee for
helpful and constructive comments concerning the intrinsic reddening 
in 3C109. SWA and ACF thank the Royal Society for support.

\end{document}